\documentstyle[12pt]{article}
\begin{document}
\begin{center}
\huge {\bf Natural boundaries for the Smoluchowski equation and
affiliated diffusion processes}
\vskip2cm
\large {\bf Philippe Blanchard and Piotr Garbaczewski\footnote{permanent and
current address:Institute of Theoretical Physics,University of Wroc\l aw,PL-50
209 Wroc\l aw,Poland;partially supported by the KBN grant No 20060 91 01}}
\vskip1,5cm
\normalsize {\em  BiBoS,Fakult\"{a}t der Physik,Universit\"{a}t
Bielefeld,D-33615 Bielefeld,Germany}
\end{center}
\thispagestyle{empty}
\vskip2cm
\begin{abstract}
The Schr\"{o}dinger problem of deducing the microscopic dynamics from
the input-output statistics data is known to admit a solution in terms
of Markov diffusions.The uniqueness of solution is found linked to the
natural boundaries respected by the underlying random motion.By choosing
a reference Smoluchowski diffusion,we automatically fix the Feynman-Kac
potential and the field of local accelerations it induces.We generate
the family of affiliated diffusions with the same local dynamics,but
different inaccessible boundaries on finite,semi-infinite and infinite
domains.For each diffusion process a unique Feynman-Kac kernel is
obtained by the constrained (Dirichlet boundary data) Wiener path
integration.As a by-product of the discussion,we give an overview
of the problem of inaccessible boundaries for the diffusion,and bring
together (sometimes viewed from  unexpected angles) results which
are little  known,and  dispersed in publications from scarcely
communicating areas of mathematics and physics
\end{abstract}
\newpage
\section{The Schr\"{o}dinger problem:microscopic dynamics from the
input-output statistics}

According to M.Kac$^{[1]}$ any kind of time developement (be it
deterministic or essentially probabilistic),which is analyzable in terms
of probability,deserves the name of the stochastic process.

Given a dynamical law of motion (for a particle as example),in many
cases one can associate with it (compute or approximate the observed
frequency data) a probability distribution and various mean values.In
fact,it is well known that inequivalent finite difference random motion
problems may give rise to the same continuous approximant (e.g. the
diffusion  equation representation of discrete processes).As well,in the
study of nonlinear dynamical systems$^{[2]}$ ,given almost any (for the
purposes of our discussion ,basically one-dimensional)probability
density,it is possible to construct an infinite number of deterministic
finite difference equations,whose iterates are chaotic and which give
rise to this a priori prescribed density.

The \it inverse operation \rm of deducing the detailed (possibly
individual,microscopic) dynamics,which either implies or is
consistent with the given probability distribution (and eventually with
its own time evolution) cannot thus have a unique solution.

If we disregard the detailed nature (like its chaotic ,jump
process,random walk,phase space process with friction etc.
implementations) of the given process,it appears$^{[3,4]}$ that the
standard Brownian motion and/or the broad class of Markovian diffusions
incorporating the Wiener noise input,provide satisfactory approximations
for a large variety of phenomena.It especially pertains to the explicit
modelling of any unknown in detail physical process in terms of the
input-output statistics (conditional probabilities and averages) of
random motions with a finite time of duration.

From now on ,we shall confine our attention to continuous Markov
processes,whose random variable $X(t),t\geq 0$ takes values on the real
line $R^1$,and in particular can be restricted (constrained) to remain
within the interval $\Lambda \subset R^1$,which may be finite or (semi-)
infinite but basically an open set.Its  boundaries $\partial \Lambda $
(endpoints) will be denoted $r_1,r_2 $ with $-\infty \leq r_1 < r_2 \leq
\infty $.

In the above input-output statistics context,let us invoke a
probabilistic problem,ori\- ginally due to Schr\"{o}dinger$^{[5-7]}$ :
\it given two strictly positive (on an open interval) boundary
probability distributions $\rho _0(x),\rho _T(x)$ for a process with the
time of duration $T\geq 0$.Can we uniquely identify the stochastic process
interpolating between them ? \rm

Perhaps unexpectedly in the light of our previous comments ,the answer
is known$^{[6,7]}$ to be affirmative ,if we assume the interpolating
process to be Markovian.In fact,we get here a unique Markovian
diffusion,which is specified by the joint probability distribution
$${m(A,B) = \int_A dx\int_B dy\,  m(x,y)}\eqno{(1)}$$
$$\int dy\,  m(x,y) = \rho _0(x) $$
$$ \int dx\,  m(x,y)=\rho _T(y)$$
where
$${m(x,y) = \Theta _*(x,0)\, k(x,0,y,T)\, \Theta (y,T)}\eqno{(2)}$$
and the two unknown (not necessarily Lebesgue integrable) functions
$\Theta _*(x,0),\Theta (y,T)$ come out as solutions of \it the same sign
\rm of the integral identities (1).Provided ,we have at our disposal a
bounded strictly positive integral kernel $k(x,s,y,t),0\leq s<t\leq
T$.Then:
$$\Theta _*(x,t) = \int k(0,y,x,t)\Theta _*(y,0)dy \eqno (3) $$
$$\Theta (x,s)=\int k(s,x,y,T)\Theta (y,T)dy$$
and the sought for interpolation has a probability distribution $\rho
(x,t) = (\Theta _*\Theta )(x,t) ,t\in [0,T]$.The transition density
$${p(y,s,x,t) = k(y,s,x,t) \, {\Theta (x,t)\over {\Theta
(y,s)}}}\eqno{(4)}$$
with $s\leq t$ is a fundamental solution ($p\rightarrow \delta (x-y)$ as
$t\downarrow s$) of the forward Kolmogorov (i.e. Fokker-Planck) equation
with a diffusion constant $D>0$ (this choice narrows slightly the
allowed framework):
$${\partial _tp = D\triangle _x p - \nabla _x (bp)}\eqno{(5)}$$
$$\rho (x,t) = \int p(y,s,x,t) \, \rho (y,s)\, dy $$
with $\rho _0(x) = \rho (x,0)$ and the drift $b(x,t)$ given by :
$$b(x,t) =2D\, {\nabla \Theta \over {\Theta }}(x,t) \eqno (6)$$
The backward diffusion equation is solved by the same transition density
$${\partial _sp = - D\triangle _yp - b\nabla _yp}\eqno{(7)}$$
$$p=p(y,s,x,t)\, ,\, s\leq t\, ,\, b=b(y,s)$$
and we deal here with a \it unique \rm diffusion process, whose
transition density is a \it common \rm fundamental solution for both the
backward and forward Kolmogorov equations.

To understand the r\^ {o}le of the integral kernel $k(y,s,x,t)$ in
(1)-(7) let us assume that $\Theta (x,t)$ is given in the form (drifts
are gradient fields as a consequence):
$$\Theta (x,t) = \pm exp\, \Phi (x,t) \Rightarrow b(x,t) = 2D\nabla
\Phi (x,t) \eqno (9) $$
$$x \in (r_1,r_2)$$
and insert (4) to the Fokker-Planck equation (5).Then$^{[8,13,14]}$,
if  $p(y,s,x,t) $ is to solve (5),the kernel $k(y,s,x,t)$
must be  a fundamental solution of the generalised diffusion
equation:
$$\partial _t\, k = D\triangle _x\, k - {1\over {2D}}\Omega (x,t)\, k
\eqno (10) $$
$$k(y,s,x,t) \rightarrow \delta (x-y) \, \hbox{\rm  as} \, t\downarrow s$$
$$\Omega (x,t) = 2D\bigl [\partial _t \Phi \, +\, {1\over 2} ({b^2
\over {2D}}+ \nabla b)\bigr ]$$
and to guarrantee (3) ,it must display the semigroup composition
properties.

Notice that (4),(9) imply that the backward diffusion equation (7) takes
the form of the adjoint to (10):
$$\partial _s\, k = - D\triangle _y\, k + {1\over {2D}} \Omega (y,s)\, k
\eqno (11) $$
$$k= k(y,s,x,t)$$

If the process takes place in-between boundaries at infinity
$r_1=-\infty ,r_2=+\infty$ ,the standard restrictions on the auxiliary
potential $\Omega $ (Rellich class$^{[15,16]}$) and hence on the drift potential
$\Phi (x,t)$ , yield the familiar Feynman-Kac representation of the
fundamental solution $k(y,s,x,t)$ \it common \rm for (10) and (11):
$$k(y,s,x,t)= \int exp\bigl [- {1\over {2D}}\int_s^t \Omega \bigl
(X(u),u\bigr )du \bigr] \, d\mu [s,y\mid t,x] \eqno (12) $$
which integrates $exp[-(1/2D)\int_s^t \Omega (X(u),u) du]$ weighting
factors with respect to the conditional Wiener measure i.e. along all
sample paths of the Wiener process which connect $y$ with $x$ in time
$t-s$.See e.g. also Refs.17,18.More elaborate discussion is necessary
,if at least one of the boundary points is \it not \rm at infinity.

Let us  notice that the time independence of $\Omega $ is granted if
either $\Phi $ is independent of time ,or depends on time at most
linearly.Then the standard expression $exp[-H(t-s)](y,x)$ for the kernel
$k$  clearly reveals the involved semigroup properties,with
$H=-D\triangle + (1/2D)\Omega (x)$ being the essentially self adjoint
operator on its (Hilbert space) domain.

\vskip1,5cm
\section {Natural boundaries make diffusions unique}

We shall make one more step narrowing slightly the scope of our
discussion by admitting diffusions (1)-(7) whose drift fields are
time-independent:$\partial _tb(x,t)=0$ for all $x$.We know$^{[8]}$ that
both the free Brownian motion and the Brownian motion in a field of
force in the Smoluchowski approximation ,belong to this class of
processes.We know also$^{[9]}$ that the boundary value problems for the
Smoluchowski equation have a profound physical significance,albeit
the attention paid to various cases is definitely unbalanced in the
literature.It is then interesting to observe that the situation we
encounter in connection with (1)-(7) is very specific from the point of
view of Feller's$^{[9-12]}$ classification of one-dimensional
diffusions encompassing effects of the boundary data.Our case is
precisely the Feller diffusion respecting (confined between)  the \it natural
boundaries \rm .An equivalent statement is that boundary points $r_1,r_2
$ are \it inaccessible \rm barriers for the process i.e. there is \it no
\rm positive probability   that any of them can be reached from the
interior of $(r_1,r_2)$ within a \it finite \rm time for all $X(0)=x \in
(r_1,r_2)$,see e.g. Refs.10,11 and 12,chap.III.4.

In the mathematical literature$^{[10,11,18]}$ a clear distinction is
made between the backward and forward Kolmogorov equations.The backward
one defines the \it transition density \rm of the process,while the
forward (Fokker-Planck) one determines the \it probability distribution
\rm (density) of diffusion.With  a given backward equation one can
usually associate the whole family of forward (Fokker-Planck)
equations,whose explicit form reflects the particular choice of boundary
data.This fundamental distinction seemingly evaporates in our previous
discussion (1)-(11) ,but it is by no means incidental.In fact ,according
to Feller$^{[10]}$:
in order that there exists one and only one (homogeneous ;
$p(y,s,x,t)=p(t-s;y,x)$) process satisfying $-\partial _tu=D\triangle u
+ b\nabla u$ in a finite or infinite interval $r_1<x<r_2$ it is
necessary and sufficient that both boundaries are inaccessible (the
probability to reach either of them within a finite time interval must
be zero).

A general feature of the inaccessible boundary problems is that the
density of diffusion vanishes$^{[9.20]}$ at the boundaries:$\rho
(r_1)=0=\rho (r_2)$. This property is shared with ,more familiar in the
realm of the statistical physics,absorbing barrier processes.The link is
indeed very close$^{[10]}$: conventionally the absorbing barrier process
is defined on the closed interval $[R_1,R_2]$ ,however we can always
consider it on  the open set $(R_1,R_2)$.Theorem 7 of Ref.10 states that
,if the boundaries $r_1<R_1<R_2<r_2$ are inaccessible for the process
,then transition densities of the absorbing barrier process on
$(R_1,R_2)$ as $R_1 \rightarrow r_1 ,R_2 \rightarrow r_2$ converge to
the unique (!) transition density of the diffusion with  unattainable
boundaries on $(r_1,r_2)$.  It implies that locally ,the inaccessible
boundary problem , in principle can be modelled (approximated) to an
arbitrary degree of accuracy by the absorbing barrier process.

It is interesting to notice that the classification of Feller's
boundaries in the homogeneous case (time independent drifts,continuous
but not necessarily bounded in the interval $(r_1,r_2)$)    follows from
investigating the Lebesgue integrability of the Hille functions (see
Refs.9,11,20 for more details) :
$$L_1 (x)= exp\bigl [-{1\over D}\int_{x_0}^x \, b(y)\, dy \bigr ]=
exp\bigl [-2[\Phi (x) -\Phi (x_0)]\bigr ]  \eqno (13) $$
$$L_2(x) = L_1 (x) \int_{x_0}^x {dz \over {L_1(z)}}=exp[-2\Phi (x)]\,
\int_{x_0}^xexp[2\Phi (z)]\, dz $$
where $x\in (r_1,r_2)$ and we have used (9) in the above.Apparently ,the
$\Phi (x_0)$ contribution in $L_1(x)$ is irrelevant,and the
integrability of $exp[\pm \Phi (x)]$ matters here.

If $L_1(x)$ is \it not \rm Lebesgue integrable on $[x_0,R]$,where $R=r_1$
or $r_2$,then $R$  stands for the  \it natural repulsive boundary of the
diffusion.If $L_1(x)$ is integrable but $L_2(x)$ is \it not \rm ,then
$R$ is a \it natural attractive \rm boundary for the process.both being
inaccesible .As indicated in Ref.9 there is no universally established
terminology
and a certain  discrimination between Feller's and the Gihman-Skorohod
definition is possible,albeit without consequences for our  discussion.

Following Ref.12 let us denote $P_x[\tau _R <\infty ]$ a probability
that a process originating from $x\in (r_1,r_2)$ would hit the point $R$
at the moment $\tau _R$ for the first time.Then, the inaccessibility of
boundaries can be expressed by the statements:
$$\hbox {\rm right boundary} \, ,\, \forall x<R \, ,\, P_x[\tau
_R<\infty]=0 $$
$$\hbox {\rm left boundary} \, ,\, \forall x>R\, ,\, P_x[\tau
_R<\infty]=0$$
For the natural boundary ,$R$ is inaccessible from the interior of
$(r_1,r_2)$ and the interior of $(r_1,r_2)$ is inaccessible from
$R$.Following this terminology: (i) An inaccessible boundary is called
\it attractive \rm ,if for any $\epsilon >0$ there exists $\delta >0$
such that $P_x[lim_{t\rightarrow \infty} X(t)=R]>1-\epsilon$ for all
$x\in (R,R+\delta)$ in case of the left ,while $x\in (R-\delta ,R)$ in
case of the right boundary.
(ii) An inaccessible (left) boundary  is called repelling ,if for any
$x>R$ and $y<x$ we have $P_y[\tau _x <\infty] = 1$.

{\bf Remark 1}: The standard (unrestricted ) Brownian motion on $R^1$ is
the most obvious example of diffusion with natural boundaries.It is not
quite trival to construct explicit examples ,if one of the boundaries
is not at infinity.The classic example of  diffusion on the half-line
with natural boundaries at $0$ and $+\infty $ is provided by the so
called Bessel process$^{[11,21]}$ ,with the diffusion (backward)
generator $L_a=\triangle _x+ {{1+2a}\over x}\nabla _x $ (we absorb the
diffusion constant $D$ in the rescaled time parameter).The point $r_1=0$
is never reached with probability one if $a\geq 0$.
In case of $a=0$ ,the transition density reads$^{[21]}$:
$${p(t;x_0,x)=const {x\over {2t}}\, exp\bigl [{-(x^2+x_0^2)\over
{4t}}\bigr ]\, I_0\bigl ({{x x_0}\over {2t}}\bigr )}  \eqno (14) $$
where the modified Bessel function (Ref.22,chap.7) is given by
$I_0(\alpha ) = \sum_{j=0}^{\infty} \alpha ^{2j} /[2^{2j}(j!)^2]$.
Another (less explicit, as given in the form of estimates for the
transition density) example pertains$^{[23]}$ to the diffusion equation
with the one dimensional harmonic oscillator potential on the half-line
$x\geq a>0$ .The related constrained path integrals are considered in
Refs.24,25.

As mentioned before ,diffusions with inaccessible barriers might have
drifts which are unbounded on $(r_1,r_2)$.Hence ,our discussion
definitely falls into the framework of diffusion processes with
singular drift fields$^{[26,27]}$,see also Refs.28-33,which is \it not
\rm covered by standard monographs on stochastic processes.Particularly
illuminating in this respect  is the analysis of Ref.26 where for quite
general diffusions, the unattainability of nodal sets (on which the
probability density vanishes) in a finite time was demonstrated ,in the
sense that $P_x[\tau _R=\infty ]=1$.
This crucial property (valid for diffusions with natural boundaries as
well) allows to extend the theory of stochastic differential equations
and integrals to diffusions , whose drifts show up a \it bad \rm
(unboundedness or divergence to infinity) behaviour when approaching the
boundaries.

We skip the standard details concerning the probability space
,filtration and the process adapted to this filtration (see however
Refs.22,34,35) and notice that a continuous random process
$X(t),t\in [0,T]$ with a probability measure $P$ is called a process of
the diffusion type if its drift $b(x)$ obeys:
$$P\bigr [\int_0^T \mid b(X(t))\mid \, dt < \infty \bigr ]=1  \eqno (15)
$$
and ,given the standard Wiener process (Brownian motion) $W(t)$ ,the
integral identity ($D$ constant and positive)
holds true P-almost surely (except possibly on sets of P-measure
zero).It means that $W(t)=(1/\sqrt {2D} [X(t) - \int_0^t b(X(s))ds]$ is
a standard Wiener process with respect to the probability measure $P$
of the process $X(t)$.

For diffusions with natural boundaries , we remain within the regularity
interval of $b(X(t))$ for all (finite) times ,and (15) apparently is
valid.Therefore,the standard rules of the stochastic It\^ {o}
calculus$^{[20]}$ can be adopted to relate the Fokker-Planck  equation
(7) with the natural boundaries to the diffusion process $X(t)$,
which$^{[35]}$ "admits the stochastic differential"
$$dX(t) =b\bigl (X(t)\bigr )dt + \sqrt {2D} dW(t)  \eqno (17) $$
$$ X(0)=x_0 \, ,\, t\in [0,T]$$
for all (finite ) times.The weak (in view of assigning the density $\rho
_0(x)$  to the random variable $X(0)$)  solution of (17) is thus well defined.

For stochastic differential equations of the form (17) ,the explicit
Wiener noise input ,because of (9) implies that irrespective of whether
natural boundaries are at infinity or not ,the
Cameron-Martin-Girsanov$^{[37-39]}$ method of measure substitutions
which parallel transformations of drifts ,is applicable.Even though the
drifts are generally unbounded on $(r_1,r_2)$ and the original
theory$^{[37-39]}$ is essentially based on the boundedness demand.
It is basically due to the fact that the probabilistic Cameron-Martin
formula relating the probability measure $P_X$ of $X(t)$ with the Wiener
meaasure $P_W$ (strictly speaking it is the Radon-Nikodym derivative of
one measure with respect to another) reduces to the familiar Feynman-Kac
formula$^{[17,18,8,14,30,36]}$ (with the multiplicative normalisation).
The problem of the existence of the Radon-Nikodym derivative (and this
of the absolute continuity of $P_X$ with respect to $P_W$ ,which implies
that
sets of $P_W$-measure zero are  of  $P_X$-measure zero as well) is
then replaced by the standard functional analytic problem$^{[15,16]}$
of representing the semigroup operator kernel via the Feynman-Kac
integral with respect to the conditional Wiener measure.

The Feynman-Kac formula is casually viewed to encompass the unrestricted
(the whole of $R^n$ ) motions,however it is known to be \it localizable
\rm ,and its validity extends also to finite and semi-infinite subsets
of $R^1$ ($R^n$more generally) as demonstrated in the context of the
statistical mechanics of continuous quantum
systems$^{[16,23,40-44]}$.More specifically,it refers to the Dirichlet
boundary conditions for self-adjoint Hamiltonians,which ensure their
essential self-adjointness (to yield the Trotter formula ).

{\bf Remark 2} :It is perhaps worthwhile to say few words about the
situation when the process , in principle can reach or cross the boundary
(nodal surface of the probability density in higher dimensions) in a
finite time.As before we limit our discussion to the stationary Markov
diffusion process and assume that the drift $b$ is a gradient $\sim
\nabla \rho /\rho $.If $\rho = exp\Phi $ where $\rho ^{1/2}$ is \it not
\rm an element of the Sobolev space
$H^1_{loc}(R^d)$ (i.e. is not integrable on bounded sets, with its first
derivative),then it is known$^{[49]}$ that  the process can reach or
cross  the set $N_+=[x\in R^d \mid \rho  =0]$
It is possible$^{[63]}$ to formulate a useful criterion (easy to generalize to
higher dimensions) for a "tunelling" (transmission) through a chosen point
in $R^1$,let it be the origin $x=0$.
Let us take $b \sim \nabla \Theta /\Theta$.The dynamics is
given by the energy form on $L^2(R,dx)$ i.e. $E[f_1,f_2)=\int \nabla f_1
\nabla f_2 \rho (x)dx $ where obviously $\rho = \Theta \Theta
_*=\Theta ^2$,compare e.g. (1)-(7).
If $\rho (x) \leq A\mid x\mid$ in the close vicinity
of the origin ,at least on one side ,then there is \it no \rm tunelling  through $x=0$ .The
process cannot cross this point ,but may be absorbed or reflected .
One needs a bit stronger restriction to prevent the diffusion from hitting
the node :non-transmission and non-hitting are not equivalent
concepts,although  there is no communication between diffusions on the positive
and negative semi-axis respectively.
On the other hand ,if $\rho (x) \geq A{\mid x\mid }^{\alpha }$ with
$0<\alpha <1$ in a neighbourhood of the origin,then there is a particle
transmission (tunelling) through $x=0$.As observed in Ref.63 ,according
to Feller's classification of boundaries,$0$ stands for the regular
boundary in this case.

\vskip1,5cm
\section{Hydrodynamic representation:local conservation laws and the
Newtonian dynamics in mean}

Let us emphasize the importance of (17), and of  the It\^{o} differential
formula induced by (17) for smooth functions of the random variable
$X(t)$. Its \it first \rm consequence is that given $p(y,s,x,t)$,for any
smooth function of the random variable the forward time derivative in
the conditonal mean  can be introduced$^{[4,7,20,27]}$ (we bypass in this
way the inherent non-differentiability of sample paths of the process)
$$lim_{\triangle t\downarrow 0} {1\over {\triangle t}}\bigl [\int p(x,t,y,t+
\triangle
t)f(y,t+\triangle t)dy - f(x,t)\bigr ] = (D_+f(X(t),t)=   \eqno (18) $$
$$=  (\partial _t + b\nabla + D\triangle )f(X(t),t)$$
$$X(t)=x$$
so that the second forward derivative associates with our diffusion  the
local field of accelerations:
$$(D^2_+X)(t) =(D_+b)(X(t),t) =(\partial _tb + b\nabla b + D\triangle
b)(X(t),t)= \nabla \Omega (X(t),t) \eqno (19) $$
with the (auxiliary potential $\Omega (x,t)$ introduced before in the
formula (10).
Since we have given $\rho (x,t)$ for all $t\in [0,T]$, the notion of the
backward transition density $p_*(y,s,x,t)$ can be introduced as well
$$\rho (x,t) p_*(y,s,x,t)=p(y,s,x,t)\rho (y,s) \eqno (20)$$
which allows to define the backward derivative of the process in the
conditional mean (cf.Refs.8,45-47 for a discussion of these concepts in
case of the most traditional Brownian motion)
$$lim_{\triangle t\downarrow 0} \, {1\over {\triangle t}}\bigl [ x -
\int p_* (y,t-\triangle t,x,t)y dy \bigr ]= (D_-X)(t)=
b_*(X(t),t)=[b- 2D\nabla ln\rho ](X(t),t) \eqno (21) $$
$$(D_-f)(X(t),t) = (\partial _t + b_* \nabla - D\triangle )f(X(t),t)$$

Apparently ,the validity of (19) (cf.Refs.7,14,27 for related
considerations) extends to $(D^2_-X)(t) $ as well ,and there holds
$$(D^2_+X)(t) = (D^2_-X)(t) = \partial _tv + v\nabla v + \nabla Q =
\nabla \Omega  \eqno (22) $$
$$v(x,t)={1\over 2} (b+b_*)(x,t) \,  ,\, u(x,t)={1\over 2}(b-b_*)(x,t) =
D\nabla ln\rho (x,t)$$
$$Q(x,t) = 2D^2\, {\triangle \rho ^{1/2}\over \rho ^{1/2}}$$
Clearly ,if $b$ and $\rho $ are time-independent ,then (22) reduces to
the identity
$$v\nabla v= \nabla (\Omega -Q) \eqno (23) $$
while in case of constant (or vanishing) current velocity $v$, the
acceleration formula (22) reduces to
$$0=\nabla (\Omega -Q) \eqno (24) $$
which establishes a very restrictive relationship$^{[48-52]}$ between
the auxiliary potential $\Omega (x)$ (and hence the drift $b(x)$) and
the probability distribution $\rho (x)$ of the stationary diffusion.The
pertinent random motions have their place in the mathematically oriented
literature$^{[48-53]}$.

Let us notice that (22) allows to transform the Fokker-Planck equation
(7) into the familiar continuity equation, so that the diffusion process
$X(t)$ admits a recasting in terms of the manifestly hydrodynamical
local conservation laws (we adopt here the kinetic theory lore)
$$\partial _t\rho  = - \nabla (\rho v) \eqno (25) $$
$$\partial _tv + v\nabla v = \nabla (\Omega - Q)$$
$$\rho _0(x)=\rho (x,0) \, ,\, v_0(x) =v(x,0)$$
which form a closed ( in fact,Cauchy) nonlinearly coupled system of
differential equations,strictly equivalent to the previous (7),(19).

In view of the natural boundaries (where the density $\rho (x,t)$
vanishes) ,the diffusion respects a specific ("Euclidean looking")
version of the Ehrenfest theorem$^{[14]}$:
$$E[\nabla Q]=0 \Rightarrow  \eqno (26) $$
$${d^2\over {dt^2}}E[X(t)]={d\over {dt}} E[v(X(t),t)]=E[(\partial _tv
+v\nabla v)(X(t,t)]=E[\nabla \Omega (X(t),t)]$$
Notice that the auxiliary potential of the form $\Omega =2Q-V$ where V
is  any Rellich class representative, defines drifts of Nelson's
diffusions$^{[14,47]}$ for which $E[\nabla Q]=0 \Rightarrow E[\nabla
\Omega ]=- E[\nabla V]$ i.e. the "standard looking" form of the second
Newton law in the mean arises.

At this point it seems instructive to comment on the essentially
hydrodynamical features (compressible fluid/gas case)of the problem
(25),where the "pressure" term $\nabla Q$ is quite annoying from the
traditional kinetic theory perspective$^{[54,55]}$.Although (25) has a
conspicuous Euler form ,one should notice that if the starting point of
our discussion would be a typical Smoluchowski diffusion$^{[8]}$
(7),(17) whose drift is given by the Stokes formula (i.e. is
proportional to the external force $F=-\nabla V$ acting on diffusing
molecules),then its external force factor is precisely the one retained
from the original Kramers phase-space formulation$^{[3,4,9]}$ of the high
friction affected random motion.
In the Euler description  of fluids and gases , the very same force
which is present in the Kramers (or Boltzmann in the traditional
discussion) equation ,should reappear on the right-hand-side of the local
conservation law (momentum balance formula) (25)  .Except for the
harmonic oscillator example ,in view of (10) it is generally \it not \rm the
case in application to diffusion processes.

Following the hydrodynamic tradition let us analyze the issue in more
detail.We consider a reference volume (control interval) $[\alpha ,\beta
]$ in $R^1$ (or $\Lambda \subset R^1$ ),which at time $t\in [0,T]$
comprises a certain fraction of particles (fluid constituents) ,for an
instant of course.Since we might deal with a flow (proportional to the
current velocity $v(x,t)$,(22))  the time rate  of particles loss by the
volume $[\alpha,\beta ]$ at time $t$ ,is equal to the flow outgoing
through the boundaries i.e.
$$-\partial _t \int_{\alpha }^{\beta }\rho (x,t)dx = \rho (\beta
,t)v(\beta ,t) - \rho (\alpha ,t)v(\alpha ,t) \eqno (27)$$
which is a consequence of the continuity equation.To analyze the
momentum balance,let us allow$^{[56]}$ for an infinitesimal deformation
of the boundaries of $[\alpha ,\beta ]$ to have entirely compensated the
mass (particle) loss (27)
$$[\alpha ,\beta ] \rightarrow [\alpha +v(\alpha ,t)\triangle t,\beta
+v(\beta ,t)\triangle t]  $$
Effectively, we pass then to the locally co-moving frame .It implies
$$lim_{\triangle t\downarrow 0} {1\over {\triangle t}}\bigl [\int
_{\alpha +v_{\alpha }\triangle t}^{\beta +v_{\beta}\triangle t} \rho
(x,t+\triangle t)dx - \int_{\alpha }^{\beta } \rho (x,t)dx\bigr ] =
\eqno (28) $$
$$= lim_{\triangle t \downarrow 0}{1\over {\triangle t}}\bigl [\int_{\alpha +
v_{\alpha
}\triangle t}^{\alpha } \rho (x,t)dx + \triangle t\int_{\alpha }^{\beta
}(\partial _t\rho ) dx + \int_{\beta }^{\beta +v_{\beta }\triangle t}
\rho (x,t) dx\bigr ]=0$$
Let us investigate what happens to the local flows $(\rho v)(x,t)$ , if
we proceed in the same way (leading terms only are retained):
$$\int_{\alpha +v_{\alpha }\triangle t}^{\beta +v_{\beta }\triangle t}
(\rho v)(x,t+\triangle t)dx - \int_{\alpha }^{\beta }(\rho v)(x,t) dt
\sim  \eqno (29) $$
$$\sim -(\rho v^2)(\alpha ,t)\triangle t + (\rho v^2)(\beta ,t)\triangle
t +\triangle t \int_{\alpha }^{\beta }[\partial _t (\rho v)]dx$$
Because of (25) we have
$$\partial _t(\rho v)=-\nabla (\rho v^2) +\rho \nabla (\Omega -Q) \eqno
(30) $$
and the rate of change of momentum associated with the control volume
$[\alpha ,\beta ]$ is
$$lim_{\triangle t\downarrow 0}{1\over {\triangle t}}\bigl [\int_{\alpha +
v_{\alpha }\triangle
t}^{\beta +v_{\beta }\triangle t} (\rho v)(x,t+\triangle t)-
\int_{\alpha }^{\beta } (\rho v)(x,t)\bigr ]= \int_{\alpha }^{\beta }
\rho \nabla (\Omega -Q) dx   \eqno (31) $$
However$^{[45]}$
$$\nabla Q = {{\nabla P}\over \rho }  \eqno (32) $$
$$P=D^2\rho \triangle ln \rho $$
and consequently
$$\int_{\alpha }^{\beta }\rho \, \nabla (\Omega -Q)dx = \int_{\alpha }^{\beta
} \rho \nabla \Omega dx - \int_{\alpha }^{\beta } \nabla P dx =  \eqno
(33) $$
$$=E[\nabla \Omega ]_{\alpha }^{\beta } + P(\alpha ,t) - P(\beta ,t) $$
Clearly ,$\nabla \Omega $ refers to the Euler-type volume force , while
$\nabla Q$ (or more correctly , $P$) refers to the "pressure" effects
entirely due to the particle transfer rate through the boundaries of the
considered volume.The latter property can be consistently attributted to
the Wiener noise proper:it sends particles away from the areas of larger
concentration.See e.g. Refs.14,45-47 and also Ref.46 for a discussion of
the Brownian recoil principle, which reverses the originally Wiener
flows.

As it appears ,the validity of the stochastic differential
representation (17) of the diffusion (5) implies the validity of the
hydrodynamical representation (25) of the process.It in turn gives a
distinguished status to the auxiliary potential $\Omega (x,t)$ of
(10)-(12).We encounter here$^{[8,14]}$ a fundamental problem of what is
to be interpreted by a physicist (observer) as the \it external force
field manifestation \rm in the diffusion process.Should it be dictated
by the drift form$^{[3,4,9,57]}$ following  Smoluchowski and Kramers ,
or rather by $\nabla \Omega $ entering the evident (albeit "Euclidean
looking") second Newton law ,respected by the diffusion ?
In the standard derivations of the Smoluchowski equation ,the
deterministic  part (force and friction terms) of the Langevin equation
is postulated.  What however,if the experimental data pertain to the
local conservation laws like (25) and (27),and there is no direct
(experimental) access to the microscopic dynamics ?

If the field of accelerations $\nabla \Omega $ is taken as the primary
defining characteristics of diffusion we deal with ,then we face the
problem of deducing all drifts ,and hence diffusions ,which give rise to
the same acceleration field,and thus form a class of \it dynamically
equivalent diffusions \rm .

\vskip1,5cm
\section{Feynman-Kac kernels on finite and semi-infinite domains with
inaccessible boundaries:dynamically equivalent diffusions}

Let us analyze the \it second consequence \rm of the unattainability  of
the boundaries ,which via (15) gives rise to (17).On the same footing as
in case of (15) ,we have satisfied another probabilistic identity:
$$P\bigr [ \int_0^T b^2\bigl (X(t)\bigr ) dt < \infty \bigr ] =1 \eqno
(34) $$
For a diffusion $X(t)$ with the differential (17) ,Theorem 6 of Ref.35
states that (34) is a sufficient and necessary condition for the
absolute continuity of the measure $P=P_X$ with respect to the Wiener
measure $P_W$.Since ,for any (Borel) set $A$ ,$P_W(A)=0$ implies
$P_X(A)=0$ , the Radon-Nikodym theorem applies$^{[34]}$ and densities of
these measures can be related.It is worthwhile to mention the
demonstration due to Fukushima$^{[49]}$ that the mutual absolute
continuity (the previous implication can be reversed) holds true for
most measures we are interested in.

In the notation  (12), the conditional Wiener measure $d\mu [s,y\mid
t,x]$ gives rise to the familiar heat kernel,if we set $\Omega =0$
identically.It in turn induces the Wiener measure $P_W$ of the set of
all sample paths,which originate from $y$ at time $s$ and terminate (can
be located) in the Borel set $A$ after time $t-s$:
$$P_W[A] = \int_A dx \int d\mu [s,y\mid t,x] =\int_A d\mu \eqno (35) $$
where ,for simplicity of notations, the $(y,t-s)$ labels are omitted and
$\int d\mu [s,y \mid t,x] $stands for the standard$^{[16]}$ path
integral expression for the heat kernel.

Having defined an It\^{o} diffusion $X(t)$ ,(5),(17) with the natural
boundaries ,we are interested in the analogous (with respect to (35))
path measure $P_X$
$$P_X[A] =\int_A dx\int d\mu _X[s,y\mid t,x] = \int _A d\mu _X  \eqno
(36) $$
The absolute continuity $P_X \ll P_W$ implies the existence of the
strictly positive Radon-Nikodym density, which we give in the
Cameron-Martin-Girsanov form$^{[34,35]}$
$${{d\mu _X}\over d\mu }[s,y\mid t,x] = exp \bigl [\int_s^t {1\over
{2D}}b(X(u))dX(u) - {1\over 2} \int_s^t {1\over {2D}} [b(X(u))]^2
du\bigr ]  \eqno (37) $$
Notice that the standard normalisation appears ,if we set $D=1/2$ which
implies $D\triangle \rightarrow {1\over 2}\triangle $ in the
Fokker-Planck equation.

On account of our demand (9) and the It\^{o} formula for $\Theta
(X(t),t)$ we have
$${1\over {2D}}\int_s^t b(X(t))dX(t) =\Theta (X(t),t) - \Theta (X(s),s)
-
\int_s^t du\, [\partial _t\Theta + {1\over 2} \nabla b ](X(u),u)  \eqno
(38) $$
so that ,apparently
$${{d\mu _X} \over d\mu }[s,y\mid t,x] =exp\bigl [ \Theta
(X(t),t)-\Theta (X(s),s)\bigr ] exp\, \bigl [-{1\over {2D}}\int_s^t
\Omega (X(u),u)du \bigr ]  \eqno (39)  $$
with $\Omega = 2D\partial _t \Theta + D\nabla b + (1/2)b^2$ introduced
before in (10) , by means of the substitution of (4) in the
Fokker-Planck equation.

In case of natural boundaries at infinity,the connection with the
Feynman-Kac formula (12) is obvious,and we have
$$P_X[A] = \int_A {{d\mu _X}\over d\mu } d\mu =\int_A dx\, \int {{d\mu
_X}\over d\mu}[s,y\mid t,x] d\mu [s,y\mid t,x]   \eqno (40)  $$
where the second integral refers to the path integration of the
Radon-Nikodym density with respect to the conditional Wiener measure,see
e.g. Refs.17,18,48-53,58.

In the context of (40) and (12) we can safely assert that the pertinent
processes ($X(t)$ and $W(t)$) have \it coinciding sets of sample paths
\rm.The stochastic process "realizes" them merely (via sampling) with a
probability distribution (frequency) different from  this for the Wiener
process $W(t)$.

The situation drastically changes ,if we wish to exploit the "likelihood
ratio" formulas (37),(39)  for diffusions confined between the
unattainable (natural) boundaries,at least one of which is \it not \rm
at infinity.
In view of the absolute continuity of $P_X$ with respect to $P_W$,we
must be able  to select a subset of Wiener paths  which coincide with
these admitted by the process  $X(t)$,except on sets  of measure zero
(both with respect to $P_X$ and $P_W$).

We face here a nontrivial problem of the existence of integral kernels
for (Schr\"{o}dinger) semigroup operators on a bounded or semi-bounded
domain.The constrained version of the formulas (40),(12) should then
integrate over a restricted set of Wiener paths:some of them must be
totally excluded ,some must "avoid" certain areas (Wiener exclusion  of
Ref.42).This problem was solved in the context of the quantum statistical
mechanics$^{[40-43,16]}$  where the Dirichlet boundary data for
self-adjoint Hamiltonian are casually associated with the completely
absorptive boundaries (cf.also Refs.30-33,36). The reason is clear ,if
one lends priority to the Brownian motion proper,since there is no
natural way for the standard Brownian motion (Wiener process) to
prohibit it from reaching or passing any conceivable boundary,except for
killing (stopping) the process,when it is going to happen.

The most transparent way towards  the localised Feynman-Kac
representation of the Dirichlet (Schr\"{o}dinger) semigroup on the
(originally$^{[36]}$ bounded) domain $\Lambda $ is by introducing the
first exit time $T_{\Lambda }$ for the Brownian path started inside
$\Lambda $ (a concrete sample path is labelled by $\omega $):
$$T_{\Lambda }(\omega ) = inf\, [t>0;X_t(\omega ) \in \Lambda ]  \eqno
(41) $$
Then ,the integral kernel of the (essentially self-adjoint on $\Lambda $,
with the Dirichlet boundary data ) Hamiltonian $H_{\Lambda }
=(-D\triangle +(1/2D)\Omega )_{\Lambda }$,is to be given by the
conditional expectation:
$$exp(-tH_{\Lambda })(s,y,t,x)=E_{y,t-s}\bigl [exp [-\int_s^t \Omega
(X_u)du]\, ;\, t<T_{\Lambda }\mid X_t =X(t)=x \bigr ]  \eqno (42) $$
which is an integration (40) restricted to these Brownian paths  which
while originating from $y\in \Lambda $ at time $s$,are conditioned to
reach $x\in \Lambda  $at time $t$,\it without crossing \rm (but
possibly touching) the boundary $\partial \Lambda $ of $\Lambda $.

Another way to write down (42) is possible ,if we define a function
$\alpha _{\Lambda }(\omega )$ on the event set (event =sample path):

$$\alpha _{\Lambda }=\cases{1,&if $X_t(\omega )\in \Lambda $ for all
$t\in [0,T]$\cr 0,&otherwise \cr }  \eqno (43) $$

Here $\alpha _{\Lambda }$ is measurable with respect to the conditional
Wiener measure,and then e.g. (35) can be replaced by the constrained
path integral$^{[23,40-43]}$:
$$P_{\Lambda }[A]  = \int_A dx\int \alpha _{\Lambda }(\omega ) d\mu
_{\omega }[s,y\mid t,x] =\int_Ad\mu _{\Lambda }  \eqno (44) $$
where $A\subset \Lambda $ and $\omega $ is the sample path label
(omitted in (35) to simplify notations).

The analysis$^{[23,40,41]}$ of special sets of Wiener measure zero is
quite illuminating at this point.Namely,the integral (44) in addition to
paths which are strictly interior to $\Lambda $,admits also paths which
\it do \rm touch the boundary $\partial \Lambda $ of $\Lambda $ for at
least one instant $t\in [0,T]$.Fortunately$^{[23,40,41]}$ ,the
$P_{\Lambda }$ measure of the set of such (unwanted) trajectories is
equal zero.

Let us now consider a diffusion $X(t)$ ,which lives in $\Lambda $ and
for which $\partial \Lambda $ is a natural boundary.Obviously \it no \rm
sample path of $X(t)$ can reach (touch) $\partial \Lambda $ in a finite
time.By the absolute continuity $P_X \ll P_W $ argument, we know that
sets of $P_W$ measure zero are the $P_X $ measure zero sets  as well.
Hence,(44) implies an apparent modification of (36):
$$P_X[A] = \int_A dx \int \alpha _{\Lambda }(\omega ) {{d\mu _X}\over
d\mu}[s,y\mid t,x] d\mu _{\omega }[s,y,t,x]  \eqno (45) $$
applicable to the diffusion $X(t)$ with the natural boundary $\partial
\Lambda $.
The Radon-Nikodym density is given by (39) and the path integral
representation (42) is apparently valid for the involved (compare e.g.
(4) again) Feynman-Kac kernel.

Except for the set of measure zero ,the process $X(t)$ is characterised
by the \it common \rm with the standard Brownian motion $W(t)$ ensemble of
sample trajectories,whose
"realisations" by $X(t)$ are appropriately (Cameron-Martin/Feynman-Kac)
weighted.

Although for each choice of the natural boundary $\partial \Lambda $
there is a unique diffusion ,which respects it , we can devise a method
of foliating the set of all considered diffusions  into \it dynamically
equivalent classes \rm .We shall
call diffusions dynamically equivalent,if they generate the same ,a
priori given field of local accelerations $b\nabla b + D\triangle
b=\nabla \Omega $ in their domain of definition.It amounts to making a
definite functional choice for $\nabla \Omega (x), x\in R^1$,and then
classifying all natural boundaries ,which are consistent with this
choice (let us emphasize that ${1\over 2}b^2 +D\nabla b =\Omega $ is to
hold true  modulo a constant).

Following Ref.8 we can always consider a given ,unrestricted in $R^1$
Smoluchowski diffusion,as the reference one.Let $\Omega (x)$ be its
auxiliary potential,and $\nabla \Omega $ the induced field of local
accelerations.What are the diffusions with natural boundaries ,which are
dynamically equivalent to this reference one ?

On purely technical grounds , the answer is simultaneously provided in
the framework of Nelson's stochastic
mechanics$^{[4,7,8,14,18,26-30,48-53,58]} $ and of Zambrini's Euclidean
quantum mechanics$^{[7,8,50-53,58,59]}$.The pertinent homogeneous
diffusions belong to the \it overlap \rm of these two theoretical
schemes and are uniquely specified by the nodal structure of stationary
solutions of the Schr\"{o}dinger-type equation:with $D$ replacing
$\hbar /2m$ in the original quantum evolution problem,and the
Schr\"{o}dinger potential being equal $\Omega (x)$ modulo an additive
(renormalisation) constant.The ground state process would correspond to
the chosen Smoluchowski diffusion.
\vskip1cm
{\bf Example}:\it The notorious (albeit exceptional) harmonic
attraction     \rm

Let us consider the Sturm-Liouville problem on $L^2(R^1)$
$$-D\triangle \psi + {{\omega ^2x^2}\over {4D}}\psi =\epsilon \psi \eqno
(46) $$
The substitutions:$\alpha ^4=\omega ^2/4D^2 ,\lambda =\epsilon /\omega
,x=\xi /\alpha $ give rise to the equivalent eigenvalue problem
$$(-{1\over 2}\triangle _{\xi} + {\xi ^2\over 2})\phi =-\lambda \phi
\eqno (47) $$
$$\phi (\xi ) = \psi ({\xi \over \alpha })=\psi (x)$$
with the well known solution (normalised relative to $x$)
$$\lambda _n=n+{1\over 2} \leftrightarrow  \epsilon _n=(n+{1\over 2})
\omega \, ,\, n=0,1,2,...              $$
$$\psi _n(x) =\phi _n(\xi ) =\bigl ({\alpha \over {2^nn!\sqrt {\pi }
}}\bigr )^{1/2} exp[-{{\xi ^2}\over 2}]\, H_n(\xi ) \eqno (48) $$
$$H_0=1\, ,\, H_1=2\xi \, ,\, H_2=2(2\xi ^2-1)\, ,\, H_3 =4\xi (2\xi
^2-3)\, ,... $$
Except for $n=0$ the solutions $\phi _n(\xi )$ are not positive definite
and change sign at nodes.We have
$$n=0 \, ,\, \psi _0(x)>0\, ,\, x\in (-\infty,+\infty )$$
$$n=1\, ,\, \psi _1(x)>0\, ,\, x\in (0,+\infty)$$
$$ \psi _1(x)<0\, ,\, x\in (-\infty,0)$$
$$n=2\, ,\, \psi _2(x)>0\, ,\, x\in (-\infty ,-1/\sqrt {2})\cup (1/\sqrt
{2},+\infty)$$
$$\psi _2 (x)<0\, ,\, x\in (-1/\sqrt {2},+1/\sqrt {2}) $$
$$n=3\, ,\, \psi _3 (x)>0\, ,\, x\in (-\sqrt {3/2},0)\cup (\sqrt
{3/2},\infty)$$
$$\psi _3(x) <0 \, ,\, x\in (-\infty ,-\sqrt {3/2})\cup
(0,\sqrt {3/2})$$
and so on.
It is convenient to continue further considerations with respect to the
rescaled $\xi =\alpha x$ variables,in view of the form $-{1\over 2}\triangle
_{\xi }+{\xi ^2\over 2} =H$ of the Hamiltonian predominantly used in the
mathematical physics literature$^{[16]}$.
To proceed in this notational convention it is enough to set
$x\rightarrow \xi $ and $D\rightarrow {1\over 2}$ in the formulas (1)-(12)
and thus utilize $b=\nabla \Theta /\Theta ,\Omega ={1\over 2}(b^2+\nabla
b),\nabla \Omega =b\nabla b +{1\over 2}\triangle b$.

Although in (1)-(12) we need $\Theta ,\Theta _*$ of the same sign,and
$\rho (x)$ to be strictly positive,we can first make a formal
identification $\Theta =\Theta _*=\phi _n,n=0,1,2,... $ and notice that
$$n=0\, ,\, b_0=-\xi \rightarrow \Omega _0={{\xi ^2}\over 2} -{1\over
2}$$
$$n=1\,   ,\, b_1={1\over \xi }-\xi \rightarrow \Omega _1= {{\xi
^2}\over 2}-{3\over 2}$$
$$n=2\, ,\, b_2={{4\xi }\over {2\xi ^2-1}}-\xi \rightarrow \Omega
_2={{\xi ^2}\over 2}-{5\over 2}$$
$$n=3\, ,\, b_3 ={1\over \xi }+ {{4\xi }\over {2\xi ^2-3}}\rightarrow
\Omega _3={{\xi ^2}\over 2}-{7\over 2}$$
Obviously $\nabla \Omega _n=\xi $ for all $n$.Irrespective of the fact
that each of $b_n,n>0$ shows singularities,the auxiliary potentials are
well defined for all $x$, and for different values of $n$ they acquire
an additive renormalisation $-\lambda _n=-(n+{1\over 2})$.

The case of $n=0$ is a canonical$^{[16,23]}$ example of the
Feynman-Kac integration,and the classic Mehler formula involves the
Cameron-Martin-Girsanov density (39) as well.

Indeed$^{[16]}$ ,the
integral kernel $[exp(-Ht)](y,x)=k(y,0,x,t)$ for $H=-{1\over 2}\triangle +
({1\over 2}x^2-{1\over 2})$ is known to be given by the formula:
$$k(y,0,x,t)=\pi ^{-1/2}(1-e^{-2t})^{-1/2}\, exp\bigl [-{{x^2-y^2}\over
2}-{{(e^{-t}y-x)^2}\over {1 - exp(-2t)}}\bigr ]  \eqno (49) $$
$$\bigl (e^{-Ht}\Theta \bigr )(x)=\int k(y,0,x,t)\Theta (y) dy $$
where the integrability property
$$\int k(y,0,x,t) exp [{{x^2-y^2}\over 2}] dy = 1  \eqno (50) $$
is simply a statement (cf.(39)) pertaining to the transition density (4)
of the homogeneous diffusion,which preserves the Gaussian distribution
$\rho (x) =(\Theta \Theta _*)(x) ={\alpha \over \sqrt {\pi }} exp(-\xi
^2) $.

The case n=1 automatically induces the (ergodic according to Ref.53)
decomposition of the diffusion process into two independent,
non-communicating components,each being confined between the natural
boundaries $(-\infty ,0)$ and $(0,\infty )$,respectively.The pertinent
processes have
the same Feynman-Kac weight in the general expression (45) for their
probability measures.Notice that we deal here with processes on the
half-line,whose drift $b_1={1\over \xi }-\xi $ has a singularity of the
Bessel type,when the diffusion is to approach the point $0$,see e.g.
Ref.60 for a related discussion.  It suggests that the construction of
the probability measure on the half-line can be accomplished by directly
starting from the Bessel process with natural boundaries at $0$ and
$\infty $.Namely ,the rescaled form of the backward Bessel generator
$$L_a={1\over 2}\triangle _\xi + {{1+2a}\over {2\xi }}\nabla _\xi  \eqno
(51) $$
with $x=\sqrt {2}\xi ,a\geq 0$,corresponds to the transition density of
the diffusion with inaccessible boundaries
$$p(t;\xi _0,\xi )=const \, {{\xi }\over t} \, exp\bigl [-{{\xi ^2 +\xi
^2_0}\over {2t}}\bigr ] \, I_a\bigl ({{\xi \xi _0 }\over t}\bigr )
\eqno (52) $$
where $I_a(\alpha )$ is the modified Bessel function.A particular choice
of $a=1/2$ i.e. $I_{1/2}(\alpha ) =\sqrt {2\over {\pi \alpha }}\sinh
{\alpha }$  provides us (in the notational convention (47)) with a
conservative diffusion whose field of drifts is $b(\xi )={1\over \xi
}$.This diffusion can be directly compared (in the sense of
Girsanov,see$^{[14,18,37,38,58,60,64]}$  to the unrestricted harmonic
diffusion considered previously:the drift transformation  from ${1\over
\xi }$ to ${1\over \xi }- \xi $ induces a corresponding transformation
of probability measures.Effectively it amounts to replacing in (45) the
restricted (to the semi-axis) conditional Wiener measure by the
conditional Bessel measure,which automatically respects the
boundaries,and next evaluating the Radon-Nikodym derivative of the
harmonic measure with respect to the Bessel measure.The corresponding
Cameron-Martin-Girsanov  (likelihood ratio ) formula can be found in
Ref.17.

The decomposition into non-communicating diffusions with natural
boundaries is characteristic of all $n>0$ solutions of (47).However all
of them induce the same local field of accelerations $\nabla \Omega (\xi
)=\xi $.

Although the existence of the Feynman-Kac kernels (and thus of the
transition densities and the diffusions themselves) is here granted ,it
is generally not easy to give analytic expressions for them.However,in
view of (44) ,the numerical simulation of each diffusion problem
encountered before ,is definitely in the reach.
Our discussion was basically one-dimensional and restricted to
stationary cases,nevertheless extensions to time-dependent
(non-stationary) processes  and to higher dimensions (much of the
outlined structure is preserved)  are available.

\vskip1,5cm

{\bf References}

1.M.Kac,J.Logan,in:Fluctuation phenomena, eds.E.W.Montroll,J.L.Lebowitz,

North-Holland,Amsterdam,1976

2.M.C.Mackey,L.Glass,From clocks to chaos:rhytms of life,Princeton Univ.

Press ,Princeton,1988

3.S.Chandrasekhar,Rev.Mod.Phys.15,(1943),1

4.E.Nelson,Dynamical theories of the Brownian motion,Princeton Univ.

Press,Princeton,1967

5.E.Schr\"{o}dinger,Ann.Inst.Henri Poincare,2,(1932),269

6.B.Jamison,Z.Wahrsch.verw.Geb.30,(1974),65

7.J.C.Zambrini,J.Math.Phys.27,(1986),3207

8.P.Garbaczewski,Phys.Lett.A 178,(1993),7

9.N.G.van Kampen,Stochastic processes in  physics and chemistry,second ed.,

North-Holland,Amsterdam,1992 ,see also$^{[61,62]}$

10.W.Feller,Trans.Am.Math.Soc.77,(1954),1

11.W.Feller,Ann.Math.55,(1952),468

12.I.Gihman,A.Skorohod,Theory of stochastic processes,vol.III,Springer,

Berlin,1979

13.R.L.Stratonovich,Select.Transl.in Math.Stat.Prob.,10 ,(1971),273

14.P.Garbaczewski,Relative Wiener noises and the Schr\"{o}dinger dynamics

in external force fields,Univ.Kaiserslautern preprint No KL-TH-93/12

15.M.Reed,B.Simon,Methods of modern mathematical physics,vol.II,

Academic Press,NY,1975

16.J.Glimm,A.Jaffe,Quantum physics:a functional integral point of view,

Springer,Berlin,1981

17.R.Carmona,in:(eds)C.Dellacherie,P.A.Meyer,M.Weil,Lecture Notes in

Math. , vol.721,Springer,Berlin,1979

18.A.Blaquiere,in: Modeling and Control of Systems,(ed) A.Blaquiere,Lect.

Notes in Control and Inform.Sci.,vol.121,Springer,Berlin,1989

19.A.T.Bharucha-Reid,Elements  of the theory of Markov processes and their

applications,McGraw-Hill,NY,1960

20.C.W.Gardiner,Handbook of stochastic methods,Springer,Berlin,1985

21.S.A.Molchanov,Theory Prob.Appl.12,(1967),310

22.S.Karlin,H.M.Taylor,A first course in stochastic processes,Academic

Press,NY ,1981

23.J.Ginibre,Some applications of functional integration in statistical

mechanics,in:Statistical mechanics and field theory,(eds) C.DeWitt,R.Stora,

Gordon and Breach,NY,1971

24.D.C .Khandekar,S.V.Lawande,Phys.Reports ,137,(1986),115

25.D.C.Khandekar,S.V.Lawande,J.Math.Phys.16,(1975),384

26.Ph.Blanchard,S.Golin,Commun.Math.Phys.109,(1987),421

27.Ph.Blanchard,Ph.Combe,W.Zheng,Mathematical and physical aspects of

stochastic mechanics,Lect.Notes in Physics,vol.281,Berlin,Springer,1987

28.W.A.Zheng,Ann.Inst.Henri Poincare,B 21,(1985),103

29.E.Carlen,Commun.Math.Phys.94,(1984),293

30.R.Carmona,in:Taniguchi Symp.PMMP,Katata 1985,Academic Press,Boston,

1987

31.M.Nagasawa,Prob.Theory Relat.Fields,82,(1989),109

32.R.Aebi,Prob.Theory Relat.Fields,96,(1993),107

33.M.Nagasawa,H.Tanaka,Z.Wahrsch.verw.Geb.68,(1985),247

34.R.S.Liptser,A.N.Shiryayev,Statistics of random processes,vol.I,

Springer,1977

35.R.S.Liptser,A.N.Shiryayev,Math.USSR Izv.6,(1972),839

36.R.Carmona,J.Lacroix,Spectral theory of random Schr\"{o}dinger operators

Birkh\"{a}user,Boston,1990

37.R.Cameron, W.Martin,Trans.Am.Math.Soc.58,(1945),184

38.I.Girsanov,Theory Prob.Appl.5,(1960),285

39.N.Bouleau,F.Hirsch,Dirichlet forms and analysis on Wiener space,de

Gruyter ,Berlin,1991

40.O.Bratteli,D.W.Robinson,Operator algebras and quantum statistical

mechanics,vol.II,Springer,Berlin,1981

41.N.Angelescu,G.Nenciu,Commun.Math.Phys.29,(1973) ,15

42.B.Simon,Adv.Math.30,(1978),268

43.B.Simon ,Functional integration and quantum physics,Academic Press,NY,

1980

44.M.Reed,B.Simon, Methods of modern mathematical physics,vol.IV,

Academic Press,NY,1978

45.P.Garbaczewski,Phys.Lett,A 162,(1992),129

46.P.Garbaczewski,J.P.Vigier,Phys.Rev.A 46,(1992),4634

47.P.Garbaczewski,Phys.Lett.A 172,(1993),208

48.M.D.Donsker,S.Varadhan,in: Functional integration and its applications,

(ed) A.M.Arthurs,Clarendon Press,Oxford,1975

49.M.Fukushima,in:Mathematics + Physics ,vol.1,(ed) L.Streit,World Scien-

tific,Singapore,1985,see also$^{[63]}$

50.S.Albeverio,R.H\o egh-Krohn,J.Math.Phys.15,(1974),1745

51.H.Ezawa,J.R.Klauder,L.A.Shepp,Ann.Phys.(NY),88,(1974),65

52.S.Albeverio,R.H\o egh-Krohn,L.Streit,J.Math.Phys.18,(1977),907

53.S.Albeverio,R.H\o egh-Krohn,Z.Wahrsch.verw.Geb.,40,(1977),1

54.A.De Masi,E.Presutti,Mathematical methods for hydrodynamic limits,

Lect.Notes Math.vol 1501,Springer,Berlin,1991

55.H.Spohn,Large scale dynamics of interacting particles,Springer,Berlin,1991

56.E.I.Verriest,D.R.Shin.Int.J.Theor.Phys.32,(1993),333

57.E.W.Larsen,Z.Schuss,Phys.Rev.B 18,(1978),2050

58.F.Guerra,Phys.Reports 77,(1981),263

59.S.Albeverio,K.Yasue,J.C.Zambrini,Ann.Inst.Henri Poincare,49,(1989),259

60.W.A.Zheng,P.A.Meyer,in:(eds) J.Azema,M.Yor,Lect.Notes Math.vol.1059,

Springer,Berlin,1984

61.W.Horsthemke,R.Lefever,Noise-induced transitions,Springer,Berlin,1984

62.H.Risken,The Fokker-Planck equation,Springer,Berlin,1989

63.S.Albeverio,M.Fukushima,W.Karwowski,L.Streit,Commun.Math.Phys.81,

(1981),501

64.H.Narnhofer,J.R.Klauder,J.Math.Phys.17,(1976),1201

\end
{document}